\def\la{\mathrel{\mathchoice {\vcenter{\offinterlineskip\halign{\hfil
$\displaystyle##$\hfil\cr<\cr\sim\cr}}}
{\vcenter{\offinterlineskip\halign{\hfil$\textstyle##$\hfil\cr<\cr\sim\cr}}}
{\vcenter{\offinterlineskip\halign{\hfil$\scriptstyle##$\hfil\cr<\cr\sim\cr}}}
{\vcenter{\offinterlineskip\halign{\hfil$\scriptscriptstyle##$\hfil\cr<\cr\sim\cr}}}}}
\def\ga{\mathrel{\mathchoice {\vcenter{\offinterlineskip\halign{\hfil
$\displaystyle##$\hfil\cr>\cr\sim\cr}}}
{\vcenter{\offinterlineskip\halign{\hfil$\textstyle##$\hfil\cr>\cr\sim\cr}}}
{\vcenter{\offinterlineskip\halign{\hfil$\scriptstyle##$\hfil\cr>\cr\sim\cr}}}
{\vcenter{\offinterlineskip\halign{\hfil$\scriptscriptstyle##$\hfil\cr>\cr\sim\cr}}}}}

\def \SAIT #1 #2 {{\em Mem.\ Soc.\ Astron.\ It.\/} {\bf #1}, #2}
\def \MESS #1 #2 {{\em The Messenger\/} {\bf #1}, #2}
\def \ASTRNACH #1 #2 {{\em Astron. Nach.\/} {\bf #1}, #2}
\def \AA #1 #2 {{\em Acta Astron.\/} {\bf #1}, #2}
\def \AAP #1 #2 {{\em Astron. Astrophys.\/} {\bf #1}, #2}
\def \AAL #1 #2 {{\em Astron. Astrophys. Lett.\/} {\bf #1}, L#2}
\def \AAR #1 #2 {{\em Astron. Astrophys. Rev.\/} {\bf #1}, #2}
\def \AAS #1 #2 {{\em Astron. Astrophys. Suppl. Ser.\/} {\bf #1}, #2}
\def \AJ #1 #2 {{\em Astron. J.\/} {\bf #1}, #2}
\def \ANNREV #1 #2 {{\em Ann. Rev. Astron. Astrophys.\/} {\bf #1}, #2}
\def \APJ #1 #2 {{\em Astrophys. J.\/} {\bf #1}, #2}
\def \APJL #1 #2 {{\em Astrophys. J. Lett.\/} {\bf #1}, L#2}
\def \APJS #1 #2 {{\em Astrophys. J. Suppl.\/} {\bf #1}, #2}
\def \APSS #1 #2 {{\em Astrophys. Space Sci.\/} {\bf #1}, #2}
\def \ASR #1 #2 {{\em Adv. Space Res.\/} {\bf #1}, #2}
\def \BAIC #1 #2 {{\em Bull. Astron. Inst. Czechosl.\/} {\bf #1}, #2}
\def \JSQRT #1 #2 {{\em J. Quant. Spectrosc. Radiat. Transfer\/} {\bf #1}, #2}
\def \MN #1 #2 {{\em Mon. Not. R. Astr. Soc.\/} {\bf #1}, #2}
\def \MEM #1 #2 {{\em Mem. R. Astr. Soc.\/} {\bf #1}, #2}
\def \PLR #1 #2 {{\em Phys. Lett. Rev.\/} {\bf #1}, #2}
\def \PASJ #1 #2 {{\em Publ. Astron. Soc. Japan\/} {\bf #1}, #2}
\def \PASP #1 #2 {{\em Publ. Astr. Soc. Pacific\/} {\bf #1}, #2}
\def \NAT #1 #2 {{\em Nature\/} {\bf #1}, #2}
\def \SSREV #1 #2 {{\em Space Sci. Rev.\/} {\bf #1}, #2}
\def \aph #1 {{\em astro-ph\/} #1}
\def \IAUC #1 {{\em IAUC} #1} 
\def\tem#1{\par\noindent
\hangindent6.5 mm\hangafter=0
\llap{#1\enspace}\ignorespaces}

\def\jref{\hang\noindent}

\documentstyle{memsait}
\input epsf.sty
\begin{opening}
\title{Be/X- RAY BINARIES}
\author{JANUSZ ZI\'O{\L}KOWSKI}
\institute{Copernicus Astronomical Center, ul. Bartycka 18, 00-716 Warsaw, Poland}
\date{} 
\end{opening}

\begin{document}

\oddpagefooter{}{}{} 
\evenpagefooter{}{}{} 
\ 
\bigskip

\begin{abstract}
A review of basic properties of Be/X-ray binaries is presented. These systems (called also hard X-ray transients), which form the most numerous class of massive X-ray binaries in the Galaxy, are composed of Be stars and neutron stars (X-ray pulsars) on wide (P$_{orb} \sim 17 - 263$ d), usually eccentric (e $\sim 0.1 - 0.9$) orbits. The systems contain two quasi-Keplerian ($\mid v_r \mid /v_{orb} \la 10^{-2}$)  discs: decretion disc around Be star and accretion disc around neutron star. Both discs are temporary: decretion disc disperses and refills on time scales $\sim$ years (dynamical evolution of the disc, formerly known as the "activity of a Be star"), while accretion disc disperses and refills on time scales $\sim$ weeks to months (which is related to the orbital motion on an eccentric orbit and, on some occasions, also to the major instabilities of the other disc). Accretion disc might be absent over a longer period of time ($\sim$ years), if the other disc is very weak or absent. The X-ray emission of Be/X-ray binaries has distinctly transient nature and is controlled by the centrifugal gate mechanism, which, in turn, is operated both by the periastron passages (Type I bursts) and by the dynamical evolution of the decretion disc (both types of bursts). The X-ray pulsars in these systems rotate at equilibrium periods (with the possible exception of the slowest pulsars). Be/X-ray binaries are excellent laboratories for investigation of both the evolutionary processes in the two discs and of the evolution of the neutron star rotation.
\end{abstract}

\section{Introduction}
At present, there are no doubts, that Be/X-ray systems (or hard X-ray transients)  dominate the population of the massive X-ray binaries. These systems, whose primaries are Be stars and secondaries are neutron stars were initially believed to be just atypical cases of massive X-ray binaries (typical cases were supergiant systems). The final Uhuru catalogue (Forman et al., 1977) listed just 1 such system (as opposed to 6 supergiant systems). Catalogue of Bradt et al. (1978), which was compilation of Uhuru, SAS-3, Ariel, Copernicus and HEAO-A sources, listed 5 Be/X-ray systems (and 8 supergiant systems). During my previous Vulcano talk on Be/X-ray binaries, 9 years ago (Zi\'o{\l}kowski, 1992), I listed 13 such binaries. Van Paradijs' (1995) catalogue contained 14 Be/X-ray systems (and 14 supergiant systems). In this review, I will present a list of 63 Be/X-ray binaries (while the number of the presently known supergiant systems is 20).

As one can see, the number of the known Be/X-ray systems is growing fast and, due to the transient nature of their emission, it is likely to continue its fast growth in the future. At the same time, the number of the known supergiant systems (in our Galaxy) is already saturated and is not expected to increase substantially,

One may notice that the situation in the field of the low mass X-ray binaries is similar. The fastest growing class of these systems is the group of soft X-ray transients (or X-ray Novae). These systems, composed typically of a black hole and a low mass optical companion, are being discovered at a high rate (again, due to the transient nature of their X-ray emission). In near future, they will, probably, dominate the population of the low mass X-ray binaries. It seems, that we have already detected almost all strong permanent X-ray sources in the Galaxy. In the future, we will detect mostly transient sources (both massive hard X-ray transients and low mass soft X-ray transients).

\section{Basic Properties}

In this section, I will list basic characteristics of Be/X-ray binaries and confront them with the properties of the other massive X-ray systems.
\medskip
\tem{(1)} The optical component is a Be star (in the other massive systems it is usually an OB supergiant).
\tem{(2)} The X-ray component is a strongly magnetized neutron star (an X-ray pulsar). In most of the supergiant systems, it is also a neutron star, but in five systems it is a black hole. No Be/black hole binary was found yet.
\tem{(3)} Orbits are elliptical with substantial eccentricities - usually e $\ga$ 0.3 (in the other systems orbits are usually circular).
\tem{(4)} The orbital periods are rather long - between 17 and 263 days (in the other systems they are almost always shorter than 10 days).
\tem{(5)} The optical components are not substantially evolved and they are substantially smaller than their Roche lobes. In the other massive systems, the optical components (supergiants) are substantially evolved (they are overluminous for their masses or undermassive for their luminosities) and they approximately fill their Roche lobes.
\tem{(6)} X-Ray emission (with a few exceptions) has distinctly transient nature with rather short active phases (a flaring behaviour). There are two types of flares, which are classified as Type I outbursts (smaller and regularly repeating) and Type II outbursts (larger and irregular). In the supergiant systems the emission is rather permanent, although it may be strongly variable.
\tem{(7)}  X-Ray emission has a hard spectrum, with typical kT  $\ga$ 15 keV (this property is shared with the other massive systems).
Because of this property (and the transient nature of their X-ray emission), the Be/X-ray binaries are also called hard X-ray transients, as opposed to the so-called soft X-ray transients (or X-ray Novae), which have typical kT $\la$ 1 keV and form a distinct group within the class of the low mass X-ray binaries.
\tem{(8)}   In all but three systems that are confirmed or probable members of the class  of Be/X-ray binaries, the neutron star is seen as an X-ray pulsar. The range of observed rotational periods covers more than four decades - from 0.03 sec to 1404 sec. In the supergiant systems the neutron star is also usually seen as an X-ray pulsar and the observed range of pulse periods is similarly wide (0.7 sec to 10000 sec).

\medskip

The list of the known Be/X-ray binaries and some of their parameters is given in Table 1. The list includes all X-ray systems containing a Be star and having at least one period (orbital or spin of the neutron star) determined.

\begin{table}
\centerline{\bf Tab. 1 $-$ Be/X-Ray Binaries}
\vspace{5mm} 
\centering
\begin{tabular}{|rcrcl|rcl|rcl|rcl|l|r|}
\hline
&&&&&&&&&&&&&&&\\
\multicolumn{5}{|c|}{Name}&\multicolumn{3}{|c|}{$P_{spin}$}&\multicolumn{3}{|c|}{$P_{orb}$}&\multicolumn{3}{|c|}{e}&\multicolumn{1}{|c|}{Sp. type}&\multicolumn{1}{|c|}{Ref}\\
&&&&&\multicolumn{3}{|c|}{[s]}&\multicolumn{3}{|c|}{[d]}&&&&&\\
&&&&&&&&&&&&&&&\\
\hline
&&&&&&&&&&&&&&&\\
SAX\hspace*{-5ex}&&J0635\hspace*{-3ex}&+&\hspace*{-2.9ex}0533&0\hspace*{-3ex}&.&\hspace*{-3ex}034&&&&&&&B0.5 IIIe&1\\
A\hspace*{-5ex}&&0538\hspace*{-3ex}&$-$&\hspace*{-3ex}7&0\hspace*{-3ex}&.&\hspace*{-3ex}069&16\hspace*{-3ex}&.&\hspace*{-3ex}7&$>$ 0\hspace*{-3ex}&.&\hspace*{-3ex}5&B2 V-IIIe&1,2\\
SMC\hspace*{-5ex}&&X\hspace*{-3ex}&$-$&\hspace*{-2.9ex}2&2\hspace*{-3ex}&.&\hspace*{-3ex}37&&&&&&&B1.5 Ve&1,3,4\\
RX\hspace*{-5ex}&&J0059.2\hspace*{-3ex}&$-$&\hspace*{-2.9ex}7138&2\hspace*{-3ex}&.&\hspace*{-3ex}763&&&&&&&B1 IIIe&1\\
AX\hspace*{-5ex}&&J0105\hspace*{-3ex}&$-$&\hspace*{-2.9ex}722&3\hspace*{-3ex}&.&\hspace*{-3ex}343&&&&&&&Be ?&1\\
4U\hspace*{-5ex}&&0115\hspace*{-3ex}&+&\hspace*{-3ex}63&3\hspace*{-3ex}&.&\hspace*{-3ex}61&24\hspace*{-3ex}&.&\hspace*{-3ex}31&0\hspace*{-3ex}&.&\hspace*{-3ex}34&B0.2 Ve&1,5\\
RX\hspace*{-5ex}&&J0502.9\hspace*{-3ex}&$-$&\hspace*{-2.9ex}6626&4\hspace*{-3ex}&.&\hspace*{-3ex}06&&&&&&&B0 IIIe&1,6\\
V\hspace*{-5ex}&&0332\hspace*{-3ex}&+&\hspace*{-3ex}53&4\hspace*{-3ex}&.&\hspace*{-3ex}37&34\hspace*{-3ex}&.&\hspace*{-3ex}25&0\hspace*{-3ex}&.&\hspace*{-3ex}31&O8.5 Ve&1,5\\
GRO\hspace*{-5ex}&&J1750\hspace*{-3ex}&$-$&\hspace*{-3ex}27&4\hspace*{-3ex}&.&\hspace*{-3ex}45&29\hspace*{-3ex}&.&\hspace*{-3ex}82&&&&Be ?&1,5\\
XTE\hspace*{-5ex}&&J0052\hspace*{-3ex}&$-$&\hspace*{-2.9ex}723&4\hspace*{-3ex}&.&\hspace*{-3ex}782&&&&&&&Be ?&4,7\\
RX\hspace*{-5ex}&&J0051.8\hspace*{-3ex}&$-$&\hspace*{-2.9ex}7231&8\hspace*{-3ex}&.&\hspace*{-3ex}9&&&&&&&Be&1\\
AX\hspace*{-5ex}&&J0049\hspace*{-3ex}&$-$&\hspace*{-2.9ex}732&9\hspace*{-3ex}&.&\hspace*{-3ex}132&&&&&&&Be ?&1\\
2S\hspace*{-5ex}&&1553\hspace*{-3ex}&$-$&\hspace*{-3ex}54&9\hspace*{-3ex}&.&\hspace*{-3ex}26&30\hspace*{-3ex}&.&\hspace*{-3ex}6&$<$ 0\hspace*{-3ex}&.&\hspace*{-3ex}09&Be ?&1,5\\
GS\hspace*{-5ex}&&0834\hspace*{-3ex}&$-$&\hspace*{-2.9ex}430&12\hspace*{-3ex}&.&\hspace*{-3ex}3&105\hspace*{-3ex}&.&\hspace*{-3ex}8&0\hspace*{-3ex}&.&\hspace*{-3ex}12&B0-2 V-IIIe&1,5\\
EXO\hspace*{-5ex}&&0531.1\hspace*{-3ex}&$-$&\hspace*{-3ex}6609&13\hspace*{-3ex}&.&\hspace*{-3ex}7&25\hspace*{-3ex}&.&\hspace*{-3ex}4&&&&Be ?&1\\
RX\hspace*{-5ex}&&J0052.1\hspace*{-3ex}&$-$&\hspace*{-2.9ex}7319&15\hspace*{-3ex}&.&\hspace*{-3ex}3&&&&&&&Be&1\\
XTE\hspace*{-5ex}&&J1946\hspace*{-3ex}&+&\hspace*{-2.9ex}274&15\hspace*{-3ex}&.&\hspace*{-3ex}8&172\hspace*{-3ex}&&&&&&Be ?&1,2\\
2S\hspace*{-5ex}&&1417\hspace*{-3ex}&$-$&\hspace*{-3ex}62&17\hspace*{-3ex}&.&\hspace*{-3ex}6&42\hspace*{-3ex}&.&\hspace*{-3ex}12&0\hspace*{-3ex}&.&\hspace*{-3ex}446&B1 Ve&1,5,6\\
GRO\hspace*{-5ex}&&J1948\hspace*{-3ex}&+&\hspace*{-3ex}32&18\hspace*{-3ex}&.&\hspace*{-3ex}76&41\hspace*{-3ex}&.&\hspace*{-3ex}7&$<$ 0\hspace*{-3ex}&.&\hspace*{-3ex}25&B0e&1,8,9\\
RX\hspace*{-5ex}&&J0117.6\hspace*{-3ex}&$-$&\hspace*{-2.9ex}7330&22\hspace*{-3ex}&.&\hspace*{-3ex}07&&&&&&&B0.5 IIIe&1\\
XTE\hspace*{-5ex}&&J1543\hspace*{-3ex}&$-$&\hspace*{-3ex}568&27\hspace*{-3ex}&.&\hspace*{-3ex}1&76\hspace*{-3ex}&.&\hspace*{-3ex}6&$<$ 0\hspace*{-3ex}&.&\hspace*{-3ex}03&B0.7 Ve&2,10\\
XTE\hspace*{-5ex}&&J0111.2\hspace*{-3ex}&$-$&\hspace*{-2.9ex}7317&30\hspace*{-3ex}&.&\hspace*{-3ex}95&&&&&&&B0-2 V-IIIe&1,11\\
RX\hspace*{-5ex}&&J0812.4\hspace*{-3ex}&$-$&\hspace*{-3ex}3114&31\hspace*{-3ex}&.&\hspace*{-3ex}89&81\hspace*{-3ex}&&&&&&B0.2 IVe&1,12,13\\
EXO\hspace*{-5ex}&&2030\hspace*{-3ex}&+&\hspace*{-3ex}37&41\hspace*{-3ex}&.&\hspace*{-3ex}7&46\hspace*{-3ex}&.&\hspace*{-3ex}03&0\hspace*{-3ex}&.&\hspace*{-3ex}41&B0e&1,2\\
XTE\hspace*{-5ex}&&J0053\hspace*{-3ex}&$-$&\hspace*{-3ex}724&46\hspace*{-3ex}&.&\hspace*{-3ex}63&139\hspace*{-3ex}&&&&&&B1 Ve&1\\
SAX\hspace*{-5ex}&&J0054.9\hspace*{-3ex}&$-$&\hspace*{-2.9ex}7226&58\hspace*{-3ex}&.&\hspace*{-3ex}97&65\hspace*{-3ex}&&&&&&B0-1 V-IIIe&1\\
Cep\hspace*{-5ex}&&X\hspace*{-3ex}&$-$&\hspace*{-2.9ex}4&66\hspace*{-3ex}&.&\hspace*{-3ex}3&&&&&&&B1e&1,2\\
RX\hspace*{-5ex}&&J0529.8\hspace*{-3ex}&$-$&\hspace*{-2.9ex}6556&69\hspace*{-3ex}&.&\hspace*{-3ex}5&&&&&&&B2 V-IIIe&1\\
RX\hspace*{-5ex}&&J0049.1\hspace*{-3ex}&$-$&\hspace*{-2.9ex}7250&74\hspace*{-3ex}&.&\hspace*{-3ex}68&&&&&&&Be&1\\
AX\hspace*{-5ex}&&J0051\hspace*{-3ex}&$-$&\hspace*{-3ex}722&90\hspace*{-3ex}&.&\hspace*{-3ex}65&120\hspace*{-3ex}&&&&&&Be&1,14\\
GRO\hspace*{-5ex}&&J1008\hspace*{-3ex}&$-$&\hspace*{-3ex}57&93\hspace*{-3ex}&.&\hspace*{-3ex}5&247\hspace*{-3ex}&.&\hspace*{-3ex}5&0\hspace*{-3ex}&.&\hspace*{-3ex}66&O9e-B1e&1,15\\
2S\hspace*{-5ex}&&1845\hspace*{-3ex}&$-$&\hspace*{-2.9ex}024&94\hspace*{-3ex}&.&\hspace*{-3ex}8&242\hspace*{-3ex}&.&\hspace*{-3ex}18&0\hspace*{-3ex}&.&\hspace*{-3ex}88&Be ?&1,16\\
RX\hspace*{-5ex}&&J0544.1\hspace*{-3ex}&$-$&\hspace*{-2.9ex}7100&96\hspace*{-3ex}&.&\hspace*{-3ex}08&&&&&&&Be&1\\
AX\hspace*{-5ex}&&J0057.4\hspace*{-3ex}&$-$&\hspace*{-2.9ex}7325&101\hspace*{-3ex}&.&\hspace*{-3ex}42&&&&&&&Be ?&4,17\\
A\hspace*{-5ex}&&0535\hspace*{-3ex}&+&\hspace*{-3ex}26&103\hspace*{-3ex}&.&\hspace*{-3ex}5&110\hspace*{-3ex}&.&\hspace*{-3ex}3&0\hspace*{-3ex}&.&\hspace*{-3ex}47&O9.7 IIIe&1,2,18\\
4U\hspace*{-5ex}&&0728\hspace*{-3ex}&$-$&\hspace*{-3ex}25&103\hspace*{-3ex}&.&\hspace*{-3ex}2&34\hspace*{-3ex}&.&\hspace*{-3ex}5&&&&O8.5 Ve&1,2\\
AX\hspace*{-5ex}&&J1820.5\hspace*{-3ex}&$-$&\hspace*{-2.9ex}1434&152\hspace*{-3ex}&.&\hspace*{-3ex}26&&&&&&&Be ?&1,19\\
RX\hspace*{-5ex}&&J0052.9\hspace*{-3ex}&$-$&\hspace*{-2.9ex}7158&169\hspace*{-3ex}&.&\hspace*{-3ex}3&&&&&&&Be&1,19\\
SAX\hspace*{-5ex}&&J1324.4\hspace*{-3ex}&$-$&\hspace*{-2.9ex}6200&170\hspace*{-3ex}&.&\hspace*{-3ex}84&&&&&&&Be&1\\
AX\hspace*{-5ex}&&J0051.6\hspace*{-3ex}&$-$&\hspace*{-2.9ex}7311&172\hspace*{-3ex}&.&\hspace*{-3ex}4&&&&&&&Be&1,20\\

&&&&&&&&&&&&&&&\\
\hline
\end{tabular}
\end{table}

\begin{table}
\centerline{\bf Tab. 1 $-$ Be/X-Ray Binaries (cont.)}
\vspace{5mm} 
\centering
\begin{tabular}{|rcrcl|rcl|rcl|rcl|l|r|}
\hline
&&&&&&&&&&&&&&&\\
\multicolumn{5}{|c|}{Name}&\multicolumn{3}{|c|}{$P_{spin}$}&\multicolumn{3}{|c|}{$P_{orb}$}&\multicolumn{3}{|c|}{e}&\multicolumn{1}{|c|}{Sp. type}&\multicolumn{1}{|c|}{Ref}\\
&&&&&\multicolumn{3}{|c|}{[s]}&\multicolumn{3}{|c|}{[d]}&&&&&\\
&&&&&&&&&&&&&&&\\
\hline
&&&&&&&&&&&&&&&\\

GRO\hspace*{-5ex}&&J2058\hspace*{-3ex}&+&\hspace*{-2.9ex}42&198\hspace*{-3ex}&&&110\hspace*{-3ex}&&&&&&Be&1\\
RX\hspace*{-5ex}&&J0440.9\hspace*{-3ex}&+&\hspace*{-
2.9ex}4431&202\hspace*{-3ex}&.&\hspace*{-3ex}5&&&&&&&B0 V-IIIe&1\\
AX\hspace*{-5ex}&&J1749.2\hspace*{-3ex}&$-$&\hspace*{-2.9ex}2725&220\hspace*{-3ex}&.&\hspace*{-3ex}38&&&&&&&Be&1,19\\
XTE\hspace*{-5ex}&&J1858\hspace*{-3ex}&+&\hspace*{-2.9ex}034&221\hspace*{-3ex}&.&\hspace*{-3ex}0&&&&&&&Be ?&1\\
GX\hspace*{-5ex}&&304\hspace*{-3ex}&$-$&\hspace*{-3ex}1&272\hspace*{-3ex}&&&132\hspace*{-3ex}&.&\hspace*{-3ex}5&$>$ 0\hspace*{-3ex}&.&\hspace*{-3ex}5&B0.7 Ve&1,2\\
AX\hspace*{-5ex}&&J0058\hspace*{-3ex}&$-$&\hspace*{-2.9ex}720&280\hspace*{-3ex}&.&\hspace*{-3ex}4&&&&&&&Be ?&1,19\\
4U\hspace*{-5ex}&&1145\hspace*{-3ex}&$-$&\hspace*{-3ex}61&292\hspace*{-3ex}&.&\hspace*{-3ex}4&187\hspace*{-3ex}&.&\hspace*{-3ex}5&$>$ 0\hspace*{-3ex}&.&\hspace*{-3ex}5&B0.2 IIIe&1,2\\
AX\hspace*{-5ex}&&J0051\hspace*{-3ex}&$-$&\hspace*{-3ex}733&323\hspace*{-3ex}&.&\hspace*{-3ex}2&1\hspace*{-3ex}&.&\hspace*{-3ex}416 ??&&&&Be&1\\
SAX\hspace*{-5ex}&&J0103.2\hspace*{-3ex}&$-$&\hspace*{-2.9ex}7209&345\hspace*{-3ex}&.&\hspace*{-3ex}2&&&&&&&O9-B1 V-IIIe&1\\
4U\hspace*{-5ex}&&2206\hspace*{-3ex}&+&\hspace*{-3ex}54&392\hspace*{-3ex}&&&9\hspace*{-3ex}&.&\hspace*{-3ex}57 ?&&&&B1e&1,21\\
A\hspace*{-5ex}&&1118\hspace*{-3ex}&$-$&\hspace*{-2.9ex}61&406\hspace*{-3ex}&.&\hspace*{-3ex}4&&&&&&&O9.5 V-IIIe&1,5\\
SAX\hspace*{-5ex}&&J1452.8\hspace*{-3ex}&$-$&\hspace*{-2.9ex}5949&437\hspace*{-3ex}&.&\hspace*{-3ex}4&&&&&&&Be ?&1\\
RX\hspace*{-5ex}&&J0101.3\hspace*{-3ex}&$-$&\hspace*{-2.9ex}7211&455\hspace*{-3ex}&&&&&&&&&Be&22\\
AX\hspace*{-5ex}&&J170006\hspace*{-3ex}&$-$&\hspace*{-2.9ex}4157&714\hspace*{-3ex}&.&\hspace*{-3ex}5&&&&&&&Be&1,19\\
AX\hspace*{-5ex}&&J0049.4\hspace*{-3ex}&$-$&\hspace*{-2.9ex}7323&755\hspace*{-3ex}&.&\hspace*{-3ex}5&&&&&&&Be&23\\
4U\hspace*{-5ex}&&0352\hspace*{-3ex}&+&\hspace*{-2.9ex}30&837\hspace*{-3ex}&.&\hspace*{-3ex}7&250\hspace*{-3ex}&.&\hspace*{-3ex}3&0\hspace*{-3ex}&.&\hspace*{-3ex}11&B0 Ve&1,24\\
RX\hspace*{-5ex}&&J1037.5\hspace*{-3ex}&$-$&\hspace*{-2.9ex}5647&862\hspace*{-3ex}&&&&&&&&&B0 Ve&1\\
SAX\hspace*{-5ex}&&J2239.3\hspace*{-3ex}&+&\hspace*{-2.9ex}6116&1247\hspace*{-3ex}&.&\hspace*{-3ex}2&262\hspace*{-3ex}&.&\hspace*{-3ex}6&&&&B0 V$-$B2 IIIe&1,25\\
RX\hspace*{-5ex}&&J0146.9\hspace*{-3ex}&+&\hspace*{-2.9ex}6121&1404\hspace*{-3ex}&.&\hspace*{-3ex}2&&&&&&&B1 Ve&1\\

1E\hspace*{-5ex}&&0236.6\hspace*{-3ex}&+&\hspace*{-3ex}6100&&& &26\hspace*{-3ex}&.&\hspace*{-3ex}49&&&&B0e&1,26\\
$\gamma$\hspace*{-5ex}&&Cas\hspace*{-3ex}&&&&&&203\hspace*{-3ex}&.&\hspace*{-3ex}59&0\hspace*{-3ex}&.&\hspace*{-3ex}26&B0.5 Ve&1,27\\
RX\hspace*{-5ex}&&J0535.0\hspace*{-3ex}&$-$&\hspace*{-2.9ex}6700&&&&241\hspace*{-3ex}&&?&&&&Be&1\\

&&&&&&&&&&&&&&&\\
\hline
\end{tabular}
\end{table}

\vspace{10mm}  

{\small NOTES:\vspace{2mm}\\
e $-$ eccentricity\\
Sp. type $-$ spectral type of the optical component \\
Ref $-$ references\\}
 
{\small REFERENCES:\vspace{2mm}\\
(1) Liu et al. 2000; (2) Okazaki and Negueruela 2001; (3) Corbet aand Marshall 2000; (4) Mereghetti 2001; (5) Bildsten et al. 1997; (6) Negueruela 1998; (7) Corbet et al. 2001; (8) Levine and Corbet 2000; (9) Negueruela et al. 2000a; (10) Marshall et al. 2000; (11) Yokogawa et al. 2000a; (12) Corbet and Peele 2000; (13) Reig et al. 2000; (14) Israel et al. 1998 (15) Negueruela and Okazaki, 2000; (16) Finger et al. 1999; (17) Torii et al. 2000; (18) Negueruela et al. 2000b; (19) Haberl and Sasaki 2000; (20) Yokogawa et al. 2000b; (21) Corbet et al. 2000; (22) Sasaki et al. 2001; (23) Ueno et al. 2000; (24) Delgado-Marti et al. 2001; (25) In't Zand et al.  2001; (26) Zamanov et al. 2001; (27) Harmanec et al. 2000.}

\vspace{8mm}

\section{Temporal Behaviour of the X-Ray Emission}

Be/X-ray binaries exhibit three types of behaviour:


(1) Type I outbursts

They last days to weeks and are connected with the periastron passages of neutron stars orbiting Be stars on alongated orbits. They repeat after the time intervals equal to the orbital periods or their multiples. The recurrency is not very strict. On some occasions, single outbursts are missing (or are unusually weak). On some others, the outbursts disappear for extended periods ($\sim$ years or decades). Almost all Be/X-ray binaries show Type I outbursts on some occasions, but the patterns are very different. Some systems (like A 0535+26) produce Type I outbursts during almost every periastron passage for a long interval of time (years or decades) and then the period of the missing outbursts follows. In some others (like 4U 0115+63 or V 0332+53) the outbursts are very rare: less than 1 \% of all periastron passages results in Type I outbursts. In the second group, the typical pattern is composed of short series of the few Type I outbursts, separated by long periods (years or decades) of inactivity, during which only Type II outbursts occur.

(2) Type II outbursts

They last several weeks and are not correlated with any particular orbital phase. These bursts are typically much stronger (by an order of magnitude or more) than Type I bursts. They occur irregularly, but the typical recurrence time is of the order of several years (although the intervals as long as two decades were also noted). Type II bursts are usually preceded by an increased activity of the Be companion (enhanced emission lines). This last fact, together with the lack of the correlation with the orbital phase, indicates that Type II bursts are connected with the activity of a Be star (while Type I bursts are clearly connected with the periastron passages).

(3) Persistent emission

Five Be/X-ray binaries are seen as permanent X-ray sources (with the emission level variable only by a factor of few). All of them are probably wide orbit ($P_{orb} \ga 200$ days) systems, containing very slow pulsars ($P_{spin}$ in the range 202 to 1404 sec). Some of these pulsars might rotate slower than at equilibrium periods (Zi\'o{\l}kowski, 2001).

\section{Why the Emission is Transient?}

In order to understand the temporal behaviour of the X-ray emission from Be/X-ray binaries, we have to invoke the, so called, accretion gate mechanism which plays a fundamental role in switching on and off the X-ray emission. Then, we have to consider the mechanisms that operate the accretion gate by controlling the supply of the external matter to the magnetosphere of the neutron star.

\subsection{The accretion gate mechanism}

The accretion gate mechanism operates due to changes in the mass flux of the matter falling on the magnetosphere of the neutron star. The size of the magnetosphere is determined by the instantaneous balance between the magnetic pressure and the dynamic pressure of the infalling matter. Therefore the magnetosphere changes its size. When the flux of the infalling matter decreases, the magnetosphere expands. When the flux increases - the magnetosphere gets squeezed. For example, the magnetosphere of a neutron star traveling along an elongated orbit around a Be star is much smaller at periastron than at apoastron. We have good reasons to believe that neutron stars in Be/X-ray binaries, similarly as many of the other X-ray pulsars, are rotating at so called equilibrium periods (Zi\'o{\l}kowski 1980, 1985; Joss and Rappaport 1984, Stella et al. 1986, Giovannelli and Zi\'o{\l}kowski 1990, and references therein). The equilibrium period is defined as a period at which the outer edge of magnetosphere rotates with the Keplerian velocity. At this period the accelerating accretion torque and the braking propeller torque should balance each other and the period, in the first approximation, should be constant. For a given magnetosphere (i.e. for a given magnetic field strength) the equilibrium period depends mainly on the mass flux of the infalling matter (because it determines the size of the magnetosphere which is assumed to corotate with the neutron star). This dependence is given as $P_{eq} \propto \dot{M}^{-3/7}$ (Davidson and Ostriker 1973, van den Heuvel 1977, Lamb 1977). Returning to our neutron star on an elongated orbit, we see that its instantaneous equilibrium period varies along the orbit as it tracks the variable mass flux. At periastron $P_{eq}$  is much smaller than at apastron. As a result, we find that in a typical situation $P_{spin} > P_{eq}$ ("slow" pulsar) at periastron but $P_{spin} < P_{eq}$  ("fast" pulsar) at apoastron. It means that at periastron the accretion gate is open and the rotation of the pulsar (neutron star) is accelerated, while at apoastron the accretion gate is closed and the pulsar is braked. The real rotation period  $P_{spin}$ will adjust itself so that the action of the accretion torque along the inner part of the orbit and the braking torque along the outer part, integrated over the full orbital cycle, will cancel each other. Therefore the statement that X-ray pulsars in Be/X-ray binaries rotate at equilibrium periods should be read as "equilibrium periods averaged over orbital cycle (or, in fact, over many orbital cycles, because of the intrinsic variability of a Be star)". One should remember, however, that along the orbit (except for the two points) the real rotation period  $P_{spin}$  substantially differs from the instantaneous value of $P_{eq}$ and, therefore, the pulsar is always, either "fast" or "slow".

The picture presented above applies directly only to those Be/X-ray binaries which exhibit (more or less regularly) Type I outbursts. However, the accretion gate mechanism plays a crucial role in all Be/X-ray binaries.

\subsection {Two states of an X-ray pulsar}

The previous paragraph leads us to the picture of two possible states of an X-ray pulsar in a binary system:

(1) Accretor

This state typically occurs when $P_{spin} > P_{eq}$. Since the velocity of the outer edge of magnetoshere is smaller then the Keplerian velocity, there is no centrifugal barrier. The matter can approach the neutron star and get accreted (accretion gate open). Due to accretion taking place, the neutron star is a strong X-ray emitter and also experiences a substantial spin-up of its rotation (the accreted matter carries a substantial angular momentum from the inner edge of the accretion disc). The characteristic properties of the accretor state are therefore: (1) "slowness" of the pulsar ($P_{spin} > P_{eq}$), (2) strong X-ray emission, (3) rapid spin-up.

(2) Propeller

This is a typical state when $P_{spin} < P_{eq}$.. In this case the velocity of the outer edge of magnetosphere is larger than the Keplerian velocity and there exist a centrifugal barrier against accretion. The matter that would like to get accreted is rather expelled by the magnetosphere acting as a propeller (llarionov and Sunyaev 1975, Davies et al. 1979). The accretion gate is closed. Only marginal amount of matter can get accreted and so the X-ray luminosity is low or undetectable. However the neutron star loses rotational energy because magnetospheric propeller has to work on the infalling matter to get it expelled and so it experiences a slow-down of its rotation. The characteristic properties of the propeller state are therefore: (1) "fastness" of the pulsar ($P_{spin} < P_{eq}$.), (2) low or undetectable X-ray emission, (3) significant spin-down.

\subsection{The Variability of the Supply of the External Matter to the Neutron Star}

From the last two paragraphs, it is clear that to operate (to open or to close) the accretion gate, one has to adjust accordingly (to increase or to decrease) the flux of the matter falling on the magnetosphere of the neutron star. In the Be/X-ray binaries, the matter in question comes from the decretion disc around the Be component (known earlier as "an envelope of a Be star"). There exist two mechanisms that might control the supply of that matter:

(1) The eccentricity of the orbit of the neutron star

The neutron star moving on an elongated orbit  around the Be star is far from the decretion disc near the apoastron and quite close to it (or may even enter it) near the periastron. This may result in the larger mass flux of the infalling matter (and the open accretion gate) near the periastron and the smaller mass flux (and the closed gate) near apoastron. This mechanism is, most likely, responsible for Type I 
outbursts, which repeat, more or less regularly, at periastron passages.

(2) The dynamical evolution of the decretion disc

Both the observational evidence and the theoretical considerations (see section 6) indicate that the decretion disc is not a stable structure and evolves dynamically on a time scale of, typically, several years. This evolution leads to cyclical dispersals and refillings of the disc. Such variability produces on some occasions large infalls of matter on the neutron star (Type II bursts). On some other occasions (e.g. empty disc) this results in so low supply of the matter to the vicinity of the neutron star that even in a system producing, rather regularly, Type I bursts, the accretion gate remains closed during periastron passages (the case of the missing Type I bursts).

\section{Accretion Discs around Neutron Stars}

Accretion discs around neutron stars in the Be/X-ray binaries have all properties of the accretion discs in other X-ray systems, except for their, possibly, transient nature. Accretion discs are, certainly, present during Type II outbursts. This might be directly inferred from the observed spin-ups of neutron stars during these bursts. The time scales of the spin-ups and their correlation with the X-ray luminosities indicate that the accreted angular momentum corresponds to the orbital angular momentum at the inner edge of the disc. The best studied case is the major February 1994 outburst of A 0535+26 (Finger et al., 1996; Bildstein et al., 1997), during which a spin-up on time scale $\sim 25$ years was seen at the peak of the outbursts. The  same outburst provided an independent evidence of the presence of an accretion disc, based on the interpretation of the QPO detected during that event. Both the frequency of the QPO (interpreted as the Keplerian frequency at the inner edge of the disc) and its correlation with the spin-up rate and with the X-ray luminosity (Finger et al., 1996; Bildstein et al., 1997) fully support the model based on an accretion disc. The accretion discs must be present also during most of Type I bursts. The system A 0535+26 demonstrated spin-ups on time scales $\sim 100$ years during many Type I bursts (Zi\'o{\l}kowski, 1985; Giovannelli and Sabau Graziati, 1992). Similar spin-ups were observed during all (2S 1845-024; Finger et al., 1999) or, at least, some Type I bursts of many other Be/X-ray systems (e.g. 4U 0115+63, GS 0834-430, 2S 1417-62, EXO 2030+375, 4U 1145-61 $-$ see Bildstein et al., 1997). Neutron stars in many Be/X-ray systems are known to experience spin-downs between the outbursts. This is in agreement with the predictions of the simple model, since they are expected to enter a propeller phase between the outbursts (see sections 4.1 and 4.2 above). For the system A 0535+26 the time scales of these spin-downs are of the order of 1000 years (Zi\'o{\l}kowski, 1985; Giovannelli and Sabau Graziati, 1992). In principle, the accretion disc may survive during the propeller phase. However, it seems that the accretion discs in the Be/X-ray binaries are not persistent. Clark et al. (1999) searched for the optical/infrared contribution from the accretion disc in A 0535+26 during the X-ray quiescent phase (propeller state). They found no significant contribution which implied that the accretion disc was either absent, or, if present, then only as a weak remnant.

Let us summarize our considerations. The accretion discs are, probably, absent during the X-ray quiescent phases. They built up on time scales of days to weeks during the periods of the enhanced infall of the matter on the magnetosphere of a neutron star (before both Type I and Type II bursts). They disperse on time scales of weeks to months, when the enhanced supply of the matter ceases.

The apparent exceptions to this picture are the five persistent systems (see section 3.3) that apparently maintain permanent accretion discs.

\section{Decretion Discs around Be Stars}

 The term "excretion disc" was introduced by B. Paczy\'nski during 1979 IAU General Assembly (Paczy\'nski, 1980). Simple, particle trajectories models of such discs were discussed by Paczy\'nski and Rudak (1980). St\c epi\'nski (1980) considered the discs around Be stars and proposed that variability of these stars is related to the changes in the sizes of the discs. The concept of the presence of two discs (one excreting and one accreting) in a Be/X-ray binary was first mentioned by Giovannelli and Zi\'o{\l}kowski (1990). Lee et al. (1991) discussed the possible existence of the outflowing viscous discs around Be stars. The next step was made when different authors (Okazaki, 1991; Papaloizu et al., 1992; Telting et al., 1994; Hanuschik et al., 1995) indicated that the one-armed instability developing in the outflowing discs may explain the, so called, V/R variability, observed in Be stars. In recent years, the outflowing viscous discs were used to describe, in details, the circumstellar matter around Be stars known earlier as "an envelope of a Be star" (Okazaki, 1997; Porter, 1999; Negueruela and Okazaki, 2000b; Okazaki, 2000; Negueruela at al., 2001; Okazaki and Negueruela, 2001). In meantime, the term "excretion" was replaced by an apparently more correct (politically) term "decretion". The modelling with the help of the viscous decretion discs appeared to be by far more succesful in describing the circumstellar matter, than earlier descriptions in terms of "equatorial winds", "expanding envelopes" or "ejected shells". In particular, the viscous disc models were able to explain the very low outflow velocities (the observed upper limits are, at most, few km/sec) and, also, to explain the V/R variability. The viscous decretion discs are very similar to the, well known, viscous accretion discs, except for the changed sign of the rate of the mass flow. Some aspects of these modellings (supply of the matter with the sufficient angular momentum to the inner edge of the disc, interaction of the stellar radiation with the matter of the disc) are not fully solved yet, but the general picture is quite convincing. The viscosity in the decretion discs is usually assumed (similarly  as for accretion discs) in the form of $\alpha$-viscosity. The discs are almost Keplerian (rotationally supported) which explains the very low values of the radial component of velocity. Nearly Keplerian discs (both inflowing and outflowing) were, since a long time, known to undergo a global one-armed oscillation instability (Kato, 1983). This instability (progressing density waves) provide a very successful explanation of V/R variability, observed both in isolated Be stars and in members of Be/X-ray systems. This phenomenon manifests itself in the form of quasi-cyclical changes of the ratio of the strengths of the V(iolet) peak to the R(ed) peak in the double profile emission lines. This variability (best seen for the H$_{\alpha}$ line) includes phases, when only one peak is visible. The time scales of the quasi-cycles range from months to years or decades. The theoretical line profiles calculated for the discs with the asymmetric matter distribution (due to progressing density waves) were found to be in a good agreement with the observed profiles (Hummel and Vrancken, 1995; Okazaki, 1996; Hummel and Hanuschik, 1997). Also the theoretical time scales calculated for the one-armed oscillation instability agreed with the observed time scales of V/R variability (Negueruela et al., 2001).

\section{The Interaction Between the Decretion Disc and the Orbiting Neutron Star}

Until recently, it was thought that this interaction works only in one direction. It is obvious that the disc can strongly influence the behaviour of the neutron star. If it sends large enough flux of the matter towards the neutron star, then the accretion gate will open and the neutron star will undergo an X-ray outbursts. It seemed that the neutron star could not significantly influence the "envelope of the Be star" (as it was termed then). This point of view changed recently. Artymowicz and Lubow (1994) investigated the truncation mechanisms for the accretion discs and found that the discs must be much smaller in the systems with large eccentricities because of the tidal/resonant truncation. The same mechanism works also for the decretion discs. The orbiting neutron star exerts a negative tidal torque on the matter of the decretion disc (removes the angular momentum). This torque diminishes the action of the viscous torque, supporting the disc, and causes that outside of the certain radius the disc cannot be supported any more (tidal truncation). This effect was investigated, for several Be/X-ray binaries, by Negueruela and Okazaki (2000a,b), and Okazaki and Negueruela (2001). They found, that the disc is usually truncated at 4:1 or 3:1 resonances (i.e. at radii for which the Keplerian period in the disc is 4 or 3 times shorter than the orbital period). This mechanism is known as tidal/resonant truncation. Due to its action, in the systems with small eccentricities, the disc is always smaller than the Roche lobe around the Be star and the flow of the matter towards the neutron star is effectively blocked. However, in the systems with large eccentricities, the disc may extend beyond the Roche lobe, during the periastron passages. The accretion on the neutron star will be then possible during these passages. Therefore, the systems in the first group (like 4U 0115+63 or V 0332+53), typically, should not produce Type I bursts, while the systems in the second group (like A 0538$-$66, A 0535+26 or GRO J1008$-$57) should produce Type I bursts during most of their periastron passages. The observations strongly support this picture.

The tidal/resonant truncation has some further consequences. Due to truncation, the matter accumulates in the outer rings of the disc. When the outer part of the disc becomes sufficiently optically thick, it may warp under the influence of the radiation of the Be star (Porter, 1998). The observations (the variability of the shape of the emission lines) tell us that the sufficiently warped disc start precessing on a time scale of weeks to months. The combined effects of the global one-armed oscillations and of the warping lead to asymmetric (elongated) configuration of the disc. Such configuration may permit an overcoming of the truncation. The, relatively, high density matter accumulated in the outer part of the disc will overflow the critical Roche lobe and fall on the neutron star leading to the Type II outburst. If the distorted disc happens to be elongated towards the periastron, it may supply the matter to the neutron star during the several periastron passages, leading to the temporary Type I outbursts in the systems, which normally do not show the bursts of this type. The Type II bursts are probably always associated with the major disturbances of the decretion disc, which frequently lead to the total disruption of the disc and the disc-less phase of the Be star. Part of the matter of the disc is accreted by the neutron star (during the Type II burst), most of it is, probably, reaccreted by the Be star. The decretion disc is then rebuilt on a time scale of, typically, a few years. The observations support the phenomenological picture presented above. In addition to the observational properties described earlier we may mention here the changes of the strength of the H$_{\alpha}$ line observed over more than 10 years period, before and after the major Type II burst in A 1118$-$61 (Coe et al., 1994). We may also add the correlation found recently for 1E 0236.6+6100/LS I+61$^o$235 system. The size of the decretion disc in this system (as measured by the strength of the H$_{\alpha}$ line) was known to vary cyclically with the periodicity of $\sim$ 4.3 years. The system was also known to exhibit radio outbursts (in addition to X-ray maxima) during the periastron passages. Zamanov and Marti (2000) found a good correlation between the strength of the radio outbursts and the strength of the H$_{\alpha}$ line as they change over the $\sim$ 1584 days modulation period.

\vspace{15mm}

\newpage
 
\section{The Comparison of Be Stars in Be/X-ray Binaries with Isolated Be Stars}

The isolated Be stars and Be stars in Be/X-ray binaries are, in many respects, quite similar. Both types have decretion discs that evolve dynamically on time scale of years to decades. In both cases, the dynamical evolution includes one-armed oscillations, V/R variability and disc-less phases. However, there are also substantial differences concerning both Be stars themselves and their decretion discs.

\subsection{Be stars}

The distribution of Be stars among different spectral subtypes was compared for both groups of Be stars by Negueruela (1998). The difference is striking. For isolated Be stars, the distribution starts at B0, rises sharply peaking at B2, declines towards B4 and then remains at approximately constant level, extending until at least A0. For Be stars in Be/X-ray binaries, the distribution starts at O8, peaks at B0 and ends at B2. The lack of the later spectral type (less massive) Be stars in binary systems may be explained with the help of the results of the evolutionary calculations by Van Bever and Vanbeveren (1997). They showed, that in scenarios involving the non-conservative mass transfer, binaries composed of a Be star and a neutron star are produced only with early type Be components. The less massive systems that would lead to binaries containing late type Be components are disrupted during the supernova explosion (the wide orbit systems are, generally, easy to disrupt).

\subsection{Decretion discs}

The decretion discs around isolated Be stars are, in many respects, similar to those in Be/X-ray binaries, as was noted earlier. The main difference is the lack of the orbiting neutron star and, therefore, the lack of the tidal/resonant truncation mechanism. As a result, the decretion discs in binaries are smaller and denser than those around the isolated Be stars (Reig et al., 1997; Negueruela and Okazaki, 2000b, Zamanov et al., 2000). As might be expected, the discs in close binaries are smaller and the discs in wide orbit binaries - larger. This is clearly seen from the strong positive correlation between the maximal observed strength of the H$_{\alpha}$ line and the orbital period of the binary, noted by Reig et al. (1997). In the few widest systems, the disc is probably only weakly perturbed by the neutron star. As a result, there are no X-ray outbursts. The low level accretion and the low level X-ray emission are maintained continuously (and the only variability is an occasional enhancement).

\section{The Pulse Period - Orbital Period Correlation for Be/X-ray Binaries (the Corbet Diagram)}

Corbet (1984) was the first to notice that there exists a strong correlation between the spin period of a neutron star in a Be/X-ray binary and the orbital period of the system. The updated version of his diagram (containing 25 systems) is shown in Fig. 1. The spread is larger than on the diagram that I have shown here nine years ago (there were only 9 points on the diagram then), but the reality of the correlation leaves no doubt. This correlation is fully to be expected, if we remember that the neutron star tries to adjust its spin period to the equilibrium value. This value increases as the flux of the matter supplied to the vicinity of the neutron star decreases ($P_{eq} \propto \dot{M}^{-3/7}$). In wide systems, with long orbital periods, the accretion is fed from the outer, low density, parts of the large (very extended) decretion disc and so the expected mass fluxes are rather low. This leads to the long spin periods.

\begin{figure}[h]
\epsfysize=10cm 
\hspace{1.1cm}\epsfbox{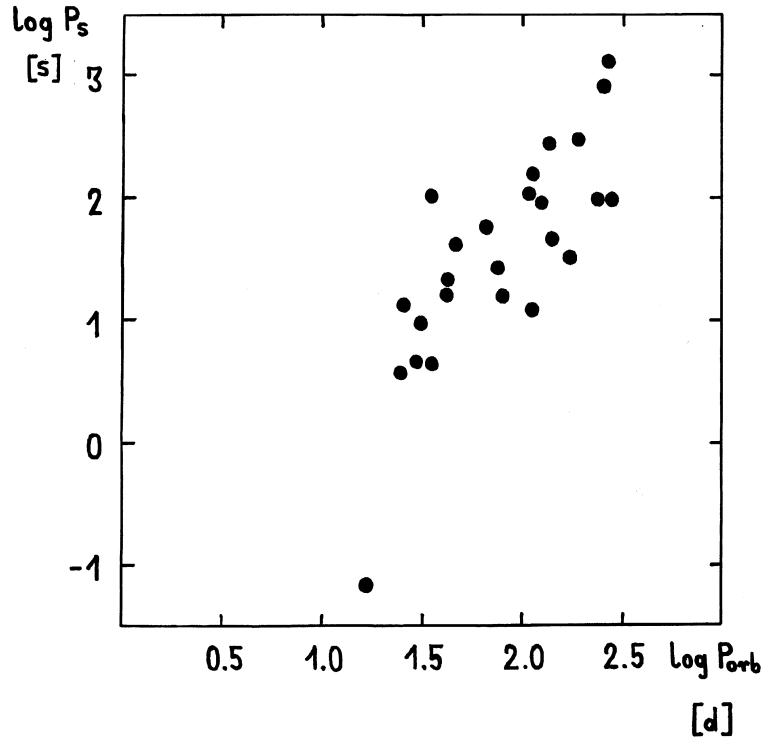} 
\caption[h]{
{\small The correlation between the spin period (in seconds) and the orbital period (in days) for Be/X-ray binaries (Corbet diagram). The diagram contains 25 systems for which both periods are presently known (the systems AX J0051-733 and 4U 2206+54 with the doubtful values of the orbital period are omitted).}}
\end{figure}

\section{The Summary}

Let us summarize the main points of our considerations.
\medskip

$\bullet$ Be/X-ray binaries (called also hard X-ray transients) form the most numerous class of massive X-ray binaries in the Galaxy (63 systems presently known as opposed to 20 supergiant systems).

$\bullet$ They are composed of Be stars and neutron stars (X-ray pulsars) on wide (P$_{orb} \sim 17 - 263$ d), usually eccentric (e $\sim 0.1 - 0.9$) orbits.

$\bullet$ X-Ray emission (with a few exceptions) has distinctly transient nature with rather short active phases (a flaring behaviour). There are two types of flares, which are classified as Type I outbursts (smaller and, more or less regularly, repeating during periastron passages over some periods of time) and Type II outbursts (larger and irregular).

$\bullet$ The optical components are not significantly evolved and they are substantially smaller than their Roche lobes.

$\bullet$  The X-ray pulsars in Be/X-ray binaries rotate at equilibrium periods (with the possible exception of the slowest pulsars). The range of the observed spin periods is very wide - from 0.03 sec to 1404 sec.

$\bullet$ The systems contain two quasi-Keplerian ($\mid v_r \mid /v_{orb} \la 10^{-2}$)  discs: a decretion disc around Be star and an accretion disc around the neutron star.
 
$\bullet$ Both discs are temporary: decretion disc disperses and refills on time scales $\sim$ years (dynamical evolution of the disc, formerly known as the "activity of a Be star"), while accretion disc disperses and refills on time scales $\sim$ weeks to months (which is related to the orbital motion on an eccentric orbit and, on some occasions, also to the major instabilities of the other disc).

$\bullet$ Accretion disc might be absent over a longer period of time ($\sim$ years), if the other disc is very weak or absent (this phenomenon manifests itself as the missing Type I bursts in the systems which normally display these bursts).

$\bullet$ The X-ray emission of Be/X-ray binaries is controlled by the centrifugal gate mechanism, which, in turn, is operated both by the periastron passages (Type I bursts) and by the dynamical evolution of the decretion disc (both types of bursts).

$\bullet$ Decretion discs are unstable to one-armed oscillation instability (which is probably responsible for the V/R variability seen in the Be stars).

$\bullet$ Decretion discs in Be/X-ray binaries are truncated by the orbiting neutron stars (the tidal/resonant truncation).

$\bullet$ Due to truncation, in the systems with small eccentricities, the flow of the matter towards the neutron star is effectively blocked during the whole orbital cycle (no Type I bursts). However, in the systems with large eccentricities, the flow is possible during the periastron passages (regular Type I bursts).
 
$\bullet$ Due to truncation, the matter accumulates in the outer rings of the decretion disc. This leads to the warping of the disc and then to its major instability (frequently to the total disruption). This mechanism is probably responsible for the Type II bursts (and for occasional Type I bursts in the systems which normally do not show them).

$\bullet$ Be/X-ray binaries are excellent laboratories for investigation of both the evolutionary processes in the two discs and of the evolution of the neutron star rotation.
\medskip

\noindent{\bf Acknowledgements}\\

This work was partially supported by the State Committee for Scientific Research grant No 2 P03C 006  19p01.
\bigskip

\noindent{\bf References}\\

\jref Artymowicz, P. and Lubow, S.H.: 1994, \APJ 421 651.

\jref Bildsten, L., Chakrabarty, D., Chiu, J. Finger, M.H., Koh, D.T., Nelson, R.W., Prince, T.A., Rubin, B.C., Scott, D.M., Stollberg, M., Vaughan, B.A., Wilson, C.A. and Wilson, R.B.: 1997, \APJS 113 367.

\jref Clark, J.S., Lyuty, V.M., Zaitseva, G.V., Larionov, V.M., Larionova, L.V., Finger, M., Tarasova, A.E., Roche, P. and Coe, M.J.: 1999, \MN 302 167.

\jref Coe, M.J., Roche, P., Everall, C., Fishman, G.J., Hagedon, K.S., Finger, M., Wilson, R.B., Buckley, D.A.H., Shrader, C., Fabregat, J., Polcaro, V.F., Giovannelli, F. and Villada, M.: 1994, \AAP 289 784.

\jref Corbet, R.H.D.: 1984, \AAP 141 91.

\jref Corbet, R.H.D. and Marshall,  F.E.: 2000, \IAUC 7402.

\jref Corbet, R.H.D., Marshall,  F.E. and Markwardt, C.B.: 2001, \IAUC 7562.

\jref Corbet, R.H.D. and Peele, A.G.: 2000, \APJL 530 33.

\jref Corbet, R., Remillard, R. and Peele, A.: 2000, \IAUC 7446.

\jref Davidson, K. and Ostriker, J.P.: 1973, \APJ 179 585.

\jref Davies, R.E., Fabian, A.C. and Pringle, J.E.: 1979, \MN 186 779.

\jref Delgado-Marti, H., Levine, A.M., Pfahl, E. and Rappaport, S.A.: 2001, \APJ 546 455.

\jref Finger, M.H., Wilson, R.B. and Harmon, B.A.: 1996, \APJ 459 288.

\jref Finger, M.H., Bildsten, L., Chakrabarty, D., Prince, T.A., Scott, D.M., Wilson, C.A., Wilson, R.B. and Nan Zhang, S.: 1999, \APJ 517 449.

\jref Giovannelli, F. and Sabau Graziati, L.:1992, \SSREV 59 1.

\jref Giovannelli, F. and Zi\'o{\l}kowski, J.: 1990, \AA 40 95.

\jref Haberl, F. and Sasaki, M.: 2000, \aph 0005226.

\jref Hanuschik, R.W., Hummel, W., Dietle, O. and Sutorios, E.: 1995, \AAP 300 163.

\jref Harmanec, P., Habuda, P., Stefl, S., Hadrava, P., Korcakova, D., Koubsky, P., Krticka, J., Kubat, J., Skoda, P., Slechta, M. and Wolf, M.: 2000, \aph 0011516.

\jref Hummel, W. and Vrancken, M.: 1995, \AAP 302 751.

\jref Hummel, W. and Hanuschik, R.W.: 1997, \AAP 320 852.

\jref Illarionov, A.F. and Sunyaev, R.A.: 1975, \AAP 39 195.

\jref In't Zand, J.J.M., Swank, J., Corbet, R.H.D. and Markwardt, C.B.: 2001, \aph 0110695.

\jref Israel, G.L., Stella, L., Campana, S., Covino, S., Ricci, D. and Oosterbroek, T.: 
1998, \IAUC 6999.

\jref Joss, P. and Rappaport, S.A.: 1984, \ANNREV 22 537.

\jref Kato, S.: 1983, \PASJ 35 249.

\jref Lamb, F.K.: 1977, in {\em Eight Texas Symposium on
Relativistic Astrophysics}, Annals New York Academy Sci. {\bf 302}, 482.

\jref Lee, U., Saio, H. and Osaki, Y.: 1991, \MN 250 432.

\jref Levine, A. and Corbet, R.: 2000, \IAUC 7523.

\jref Liu, Q.Z., van Paradijs, J. and van den Heuvel, E.P.J.: 2000, \AAS 147 25.

\jref Marshall, F.E., Takeshima, T. and in't Zand, J.: 2000, \IAUC 7363.

\jref Mereghetti, S.: 2001, in {\em Frontier Objects in Astrophysics and Particle Physics}, F. Giovannelli and G. Mannocchi, eds., Italian Physical Society, Editrice Compositori, Bologna, Italy,
 {\bf 73}, 239.

\jref Negueruela, I.: 1998, \aph 9807158.

\jref Negueruela, I., Marco, A., Speziali, R. and Israel, G.L.: 2000a, \IAUC 7541.

\jref Negueruela, I. and Okazaki, A.T.: 2000a, \aph 0011406.

\jref Negueruela, I. and Okazaki, A.T.: 2000b, \aph 0011407.

\jref Negueruela, I., Reig, P., Finger, M.H. and Roche, P.: 2000b, \aph 0002272.

\jref Negueruela, I., Okazaki, A.T., Fabregat, J., Coe, M.J., Munari, U. and Tomov, T.: 2001, \aph 0101208.

\jref Okazaki, A.T.; 1991, \PASJ 43 75.

\jref Okazaki, A.T.; 1996, \PASJ 48 305.

\jref Okazaki, A.T.; 2000, \aph 0010517.

\jref Okazaki, A.T.; 1997, in {\it The Interactions of Stars with Their Environment}, L.V. Toth, M. Kun and L. Szabados (eds.), Konkoly Observatory, Budapest, p. 407.

\jref Okazaki, A.T. and Negueruela, I.: 2001, \aph 0108037.

\jref Paczy\'nski, B.: 1980, {\it Highlights of Astronomy} {\bf 5}, 27.

\jref Paczy\'nski, B. and Rudak, B.: 1980, \AA 30 237.

\jref Papaloizu, J.C., Savonije, G.J. and Henrichs, H.F.: 1992, \AAL 265 45.

\jref Porter, J.M.: 1998, \AAP 336 966.

\jref Porter, J.M.: 1999, \AAP 348 512.

\jref Reig, P., Fabregat, J. and Coe, M.J.: 1997, \AAP 322 193. 

\jref Reig, P., Negueruela, I., Buckley, D.A.H., Coe, M.J., Fabregat, J. and Haigh, N.J.: 2000, \aph  0011305.

\jref Sasaki, M., Haberl, F., Keller, S. and Pietsch, W.: 2001, \aph 0102361.

\jref Stella, L., White, N.E. and Rosner, R.: 1986, \APJ 308 669.

\jref St\c epi\'nski, T.: 1980 \AA 30 413.

\jref Telting, J.H., Heemskerk, M.H.M., Henrichs, H.F. and Savonije, G.J.: 1994, \AAP 288 588.

\jref Torii, K., Kohmura, T., Yokogawa, J. and Koyama, K.: 2000, \IAUC 7441.

\jref Ueno, M., Yokogawa, J., Imanishi, K. and Koyama, K.: 2000, \IAUC 7442.

\jref Van Bever, J. and Vanbeveren, D.: 1997, \AAP 322 116.

\jref Van den Heuvel, E.P.J.: 1977, in {\em Eight Texas 
Symposium on Relativistic Astrophysics}, Annals New York Academy   Sci. {\bf 302}, 482.

\jref Van Paradijs, J.: 1995, in {\em X-Ray Binaries}, W.H.G. Lewin, J. van Paradijs and E.P.J. van den Heuvel, (eds.), Cambridge University Press, p. 536.

\jref Yokogawa, J., Paul, B., Ozaki, M., Nagase, F., Chakrabarty, D. And Takeshima, T.: 2000a, \APJ 539 191.

\jref Yokogawa, J., Torii, K., Imanishi, K. and Koyama, K.: 2000b, \PASJ 52 L37.

\jref Zamanov, R. and Marti, J.: 2000, \aph 0005201.

\jref Zamanov, R., Marti, J. and Marziani, P.: 2001, \aph 0110114.

\jref Zamanov, R.K., Reig, P., Marti, J., Coe, M.J., Fabregat, J., Tomov, N.A. and Valchev, T.: 2000, \aph 0012371.

\jref Zi\'o{\l}kowski, J.: 1980, in {\em Close Binary Stars: Observations and Interpretations}, M.J. Plavec, D.M. Popper and R.K. Ulrich (eds.), D. Reidel Publ. Co., Dordrecht, Holland, p. 335.

\jref Zi\'o{\l}kowski, J.: 1985, \AA 35 185.

\jref Zi\'o{\l}kowski, J.: 1992, {\it Be/X-Ray Binaries}, review presented at Vulcano Workshop 1992 (unpublished).

\jref Zi\'o{\l}kowski, J.: 2001, in {\it  Exploring the Gamma-Ray Universe}, A. Gimenez, V. Reglero and C. Winkler (eds.), (Proceedings of the Fourth Integral Workshop), ESA SP-459, p. 169.

\end{document}